# An Advanced Two-Stage Grid Connected PV System: A Fractional-Order Controller

Shah Fahad*, Nasim Ullah**, Ali Jafer Mahdi ***, Asier Ibeas****, Arman Goudarzi*‡

* College of Electrical Engineering, Zhejiang University, Hangzhou, China

**Department of Electrical Engineering, University of Technology, Nowshera, Pakistan

***Department of Electrical and Electronics Engineering, University of Kerbala, Iraq

****Escola d'Enginyeria, Universitat Autonoma de Barcelona, Barcelona, Spain

(shah.fahad072@zju.edu.cn , nasim.ullah@uotnowshera.edu.pk, ali.j.mahdi@uokerbala.edu.iq, asier.ibeas@uab.cat, agoudarzi@zju.edu.cn)

‡ Corresponding Author; Last Author, College of Electrical, Zhejiang University, Hangzhou, China

Tel: +86 137 7789 4042, agoudarzi@zju.edu.cn

**Abstract-** A fractional-order (FO) based controller for a grid-connected PV system is presented in this paper. A single phase two-stage grid-connected photovoltaic generator (PVG) is used to test the performance of the FO controller. The main objectives of the proposed controller are: (1) To regulate the output voltage of PVG at a point where the maximum power is drawn. (2) Constant DC-link voltage control. (3) Power factor control (PFC) at the inverter output with low total harmonic distortion (THD). To solve the first problem, a non-linear control method known as fractional-order back-stepping control (FOBSC) is used to regulate the output voltage of PVG. A maximum power point tracking (MPPT) technique known as perturb and observe (P&O) is used to generate a reference voltage which is suitable for tracking the maximum power generation of PVG. The generated reference is used to regulate the output voltage of PVG using FOBSC. The DC-link voltage fluctuation issue is tackled using FO based PI controller. The last objective is achieved using FOBSC to obtain maximum power factor of the grid. Lyapunov candidate function is used to verify the stability of the system. To test the performance of the proposed controller, it is compared to conventionally known Integer-order (IO) controller. Results have shown a significant improvement in THD and efficiency of the system. The proposed controller offers 0.94%, 1.43% and 1.86% lower THD in comparison with IO controller at 100%, 80% and 70% of the power generation capacity of PVG, respectively. The overall efficiency of the system for 100%, 80%, and 70% of the dynamic powers of the system is noticed to be better in case of FO controller.

*Nomenclature*

*Indexes*

| | |
|---|---|
| BSC | Back-stepping controller |
| FO | Fractional-order |
| FOBSC | Fractional-order back-stepping control |
| FOPI | Fractional-order proportional integral |
| IO | Integral-order |
| IOBS | Integral-order back-stepping |
| IV-Curve | Current-voltage curve of a photovoltaic generator |
| MPPT | Maximum power point tracking |
| P & O | Perturb and observe |
| PVG | Photovoltaic generator |
| PF | Power factor |
| RCC | Ripple co-relation control |
| THD | Total harmonic distortion |

*Variables*

| Symbol | Description |
|---|---|
| $_aD_t^\alpha$ | Fractional-order operator |
| $\alpha$ | Fractional-order operator's coefficient |
| $C_{dc}$ | DC-link capacitor |
| $C_{pv}$ | Capacitor parallel to PVG |
| $I_o$ | Boost converter inductor current |
| $K_I$ | Integral constant of PI controller |
| $K_P$ | Proportional constant of PI controller |
| $L_g$ | Grid inductor |
| $L_o$ | Boost converter inductor |
| $M_1, M_2$ | Alternate switches (1 and 2) of inverter |
| $M_3, M_4$ | Alternate switches (3 and 4) of inverter |
| $V_{dc}$ | Voltage across DC-link capacitor |
| $V_g$ | Grid voltage |
| $V_1, V_2$ | Lyapunov stability candidate functions |
| $\alpha_1$ | FOBSC tuning order 1 |
| $\alpha_2$ | FOBSC tuning order 2 |

## 1. Introduction

In the past few decades, because of the rapid increase in population and global industrialization, demand for electricity has grown substantially. Due to abundance of renewable resources in nature, solar energy is considered as one of the most promising forms of renewable energy resources. Solar energy has zero fuel cost and also it is an environmentally friendly source of energy. Much interest is being taken in designing an efficient PV system in which it has fewer grid-connectivity challenges such as grid stability, generation intermittency, power mismatch and etc. have been investigated [1, 2, 4, 38, 39].

Generally, PVGs have two major flaws. The first and most important one is the efficiency which has been calculated as low as 9-16% according to [6]. The other drawback is non-linearity of current-voltage (IV) characteristic of PV generation unit. The IV-characteristic of PVG changes with irradiance and the ambient temperature of the cells [7].

Non-linearity of the PVG unit has been one of the major concerns of researchers while studying the integration of solar energy. To tackle this problem, a technique called maximum power point tracking (MPPT) is used to extract maximum power from the PV source and deliver it to the load [5]. In this study, the P&O method is used for the implementation of MPPT and enhancing the converter efficiency. The efficiency of the implemented P&O algorithm is about 91.4% at the full irradiance condition and 95.6% at the partial cloudy conditions [5, 9].

For the operation of a PV system, a DC-DC converter is required to increase the output voltage of the PVG unit higher than the grid nominal voltage to ensure the uni-directional power flow from source to the grid. In this regard, for a grid-tied inverter configuration, a DC-DC boost converter is required to provide a constant power flow to the input of inverter [8, 10].

The power balance is essential to be maintained on the DC-link side. At the time that the power is being fed to the inverter through the DC-link, the voltage of the DC-link varies with power transfer. As the power flow-towards the DC-link increases, accordingly the voltage across it increases [10, 11, 40]. Thus, the input to the inverter changes with a change in voltage of DC-link. Therefore, a feedback control loop is required to keep the voltage across DC-link constant.

With respect to the above-mentioned reasons, an important consideration for a grid-connected PV inverter is to regulate the DC-link voltage to a level higher than grid voltage. Therefore, it is necessary to insert a filter before feeding the power to the grid in order to reduce the THD produced due to the high switching frequency of the PV inverter. As per IEEE standards, the total permissible THD of a grid-tied inverter should be less than 5% [3].

In [30], the authors claim to have presented a cost-effective grid-connected PV inverter. The THD at full load is achieved at 6.7% using an LC filter which is still above the IEEE THD grid code standard. The author in [22], has applied a hybrid technique for optimizing switching angles of the inverter for eliminating selective harmonics. A lot of work has been done in finding the appropriate mathematical solutions while the THD is reduced only to 4.61%. In [25], a control method based on virtual impedance for increasing output impedance of inverter is used to improve the system stability, where the THD could be reduced to 4.12%. In [32], a simplified DQ controller for a PV system has been presented. Although the chance of ripple attenuation using LCL filter is higher, the obtained THD was not noticeably low.

Active and reactive power control strategy has been implemented for grid-connected PV inverter by the authors in [24], in which the control system is based on dq-transformation. In [23], ripple co-relation control (RCC) is discussed which has a drawback of unstable behaviour during irradiance variation. An improved control method called hybrid RCC is proposed which mitigates the drawback of previous control. In [12], [21-26] and [33-35], the authors proposed integer order back-stepping control and integer order integral back-stepping control respectively. In [7], the author has applied a variable structure control for tuning power factor along with DC-link voltage control and maximum power point tracking. In [31], the study presented an adaptive filtering method for reducing THD by addressing double line frequency voltage ripples in voltage across dc-link. In this study, the THD has been reduced, whereas a quantitative analysis was not presented. In all the aforementioned papers, the studies were based on power quality but lack quantitative analysis. To the best of our knowledge, the fractional order back-stepping controller for a two-stage grid-connected PV inverter using L filter has never been considered. The slow and intense decay nature of fractional order derivative and integral can be further exploited to achieve a better power quality in a grid-connected PV system. A fractional order controller allows more degree of freedom as compared to an integer order controller. A simple L filter is preferred over LC and LCL filter for PV systems due to the following features: (i) simplicity in design, (ii) low cost and (iii) lack of resonant effect [29]. Furthermore, the stability proof using Lyapunov stability theory is presented. A simple L-filter has only one dynamic equation to be considered for the stability proof in Lyapunov stability theory, thereby, leads to reduction of complexity in deriving the essential equations.

In this article, the power quality enhancement has been given a priority. Reducing the THD and increasing the efficiency of the system by exploiting the slow and intense nature of the FO controller is the main contribution of this work. FO controller has a wider range of stability margins as well as the release of energy is less and slower as compared to the integer order system. Based on the previously discussed literature, fractional order back-stepping controller for PV inverters has never been discussed. This article proposes the implementation and derivation of fractional order back-stepping controller (FOBSC) for tracking the reference voltage generated by the MPPT and power factor control for sine wave grid-connected PV inverter. The P&O algorithm is used to generate a voltage reference which is traced by FOBS controller. The DC-link voltage is regulated using a fractional order PI (FOPI) controller. Finally, the current through L-filter is regulated to the phase of grid voltage using FOBSC to achieve unity power factor.

The main contributions of the study can be listed as follows:

- A fractional order back-stepping controller (FOBSC) is derived and implemented for a two-stage grid-connected PV inverter which uses an L-filter to enhance THD.
- The slow and intense nature of fractional order controller has been exploited to enhance the power quality of the PV system.
- A comparison between the proposed fractional order back-stepping controller (FOBSC) to a conventional integer order controller is presented in terms of power quality.

This paper is presented in the following manner: In section 2 mathematical model of the complete system is given. Derivation of the proposed controller is presented in section 3. Simulation results and discussions are given in section 4. Section 5 summarizes the conclusions of all the work that has presented in this research.

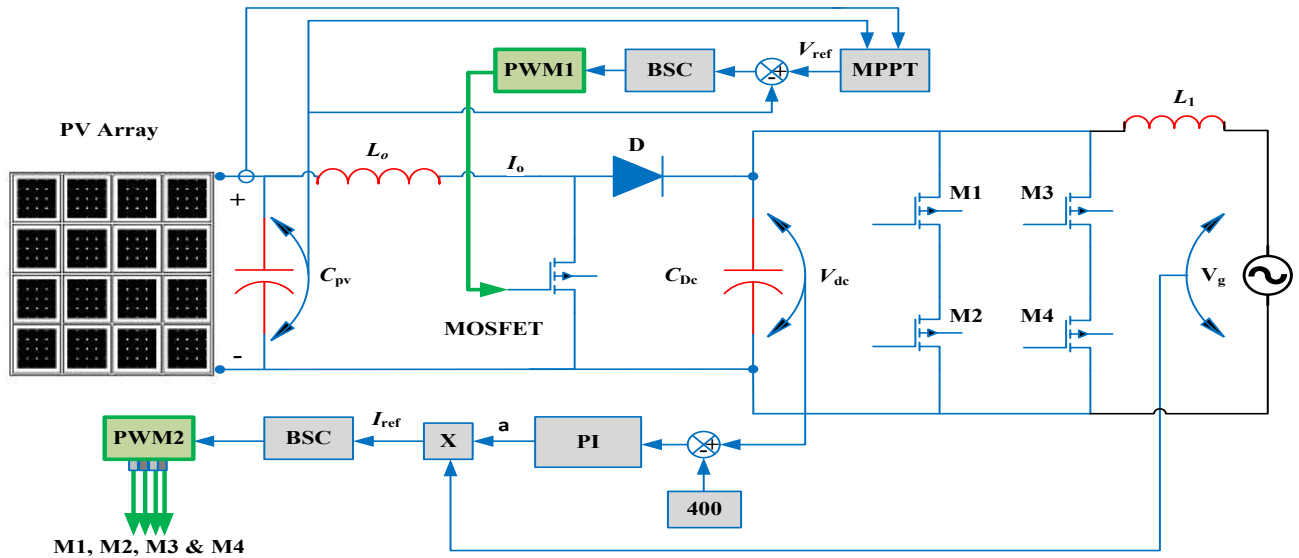

**Fig. 1:** Block diagram of the complete system

## 2. Mathematical Preliminaries and Modelling

The complete circuit diagram along with the controller layout is given in Fig. 1. The DC-link capacitor, denoted by $C_{DC}$, plays an important role of feeding a constant DC power to the inverter. For better understanding of the proposed idea, the system can be divided into two parts. From the left-hand side to the DC-link capacitor is the DC part which includes PVG and the boost converter. The second part of the system, is the

AC part in which a constant DC-link voltage is fed to an inverter to convert it to AC power. This inverted power consists of harmonic distortions due to high switching frequency mechanism of the inverter, therefore a filter is used to reduce THD and feed this power to the grid.

## 2.1 Photovoltaic Module

The PVG used in this system has an IV-Characteristic curve shown in Fig. 7a and 7b.

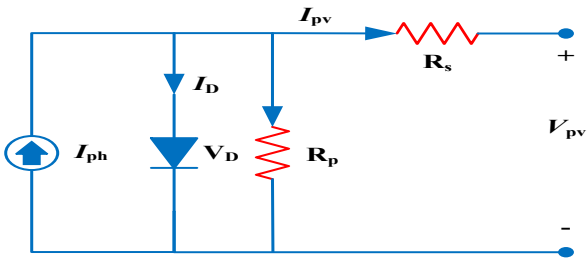

**Fig. 2:** Electrical Representation of PVG

The current generated by a PV module is presented in the following equation:

$$I_{pv} = I_{ph} - I_o \left[ exp\left(\frac{q(V+IR_s)}{N_s KTA}\right) - 1 \right] - (V + IR_s)/(R_p) \quad (1)$$

Where $q$ is electron charge, $K$ is the Boltzmann Constant, $T$ is the PV module temperature, $I_o$ is the reverse saturation current of the diode, A is the diode ideality constant, $I_{ph}$ is the light generated current a of PV cell, $R_p$ is the shunt resistance of PV cell, $R_s$ is the series resistance of the PV cell, $N_s$ is the number of the PV module connected in series and $I_{pv}$ is the output current [28].

## 2.2 DC-DC Boost Converter

A DC-DC boost converter provides an interface between PVG and inverter while allowing us to implement different control strategies including MPPT. A PV inverter requires a constant DC input voltage to operate at its best. A stable input voltage with least ripples can promise a better inverter output [37, 41]. Hence, a boost converter as shown in Fig. 3 is attached between inverter and PVG to keep the input voltage to the inverter constant. A boost converter consists of an inductor, dc-link capacitor (output capacitor), a transistor and a diode [42].

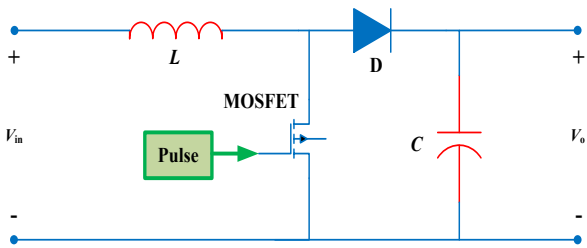

**Fig. 3:** Boost Converter

$$L_{min} \geq \frac{V_{in}(V_{out}-V_{in})}{\Delta I * V_{out} * f_s} \quad (2)$$

$$V_{out} = \frac{V_i}{1-D} \quad (3)$$

$$D = \frac{t_{on}}{T} \quad (4)$$

Where $D$ is the duty cycle, $t_{on}$ is the time for which switch is turned on, $f_s$ is the switching frequency of the boost converter, $L_{min}$ is the minimum input inductance of the boost converter. The output of the PV inverter capacitor of the boost converter is measured through the following equation:

$$C_{dc} = \frac{P_g}{\Delta V V_{dc} \omega} \quad (5)$$

Where $P_g$ is the power generated by PVG, $\Delta V$ is the voltage ripple normally kept about 10% of total voltage, $V_{dc}$ is the DC-link voltage, $\omega$ is the grid angular frequency. Any further required explanation for the classical inverter has been given in [36].

## 2.3 Mathematical Background of the Fractional-Order Operator

In this section, the mathematical preliminaries are introduced for the non-integer order calculus. Fractional-order operator is defined as $_aD_t^\alpha$

$$_aD_t^\alpha \cong D^\alpha = \begin{cases} \frac{d^\alpha}{dt^\alpha} & , R(\alpha) > 0 \\ 1 & , R(\alpha) = 0 \\ \int_a^t (d\tau)^{-\alpha} & , R(\alpha) < 0 \end{cases} \quad (6)$$

In the above equation, $\alpha$ represents fractional operator's order, while $R(\alpha)$ represents the set of real numbers. Three main definitions of general fractional operators are discussed below [16]:

**Definition 1:** The $\alpha^{th}$ order Riemann–Liouville fractional derivative and integration of a function $f(t)$ with respect to $t$ is given by:

$$_aD_t^\alpha f(t) = \frac{d^\alpha}{dt^\alpha} f(t) = \frac{1}{\Gamma(m-\alpha)} \frac{d^m}{dt^m} \int_a^t \frac{f(\tau)}{(t-\tau)^{\alpha-m+1}} d\tau \quad (7)$$

$$_aD_t^{-\alpha} f(t) = I^\alpha f(t) = \frac{1}{\Gamma(\alpha)} \int_a^t \frac{f(\tau)}{(t-\tau)^{1-\alpha}} d\tau \quad (8)$$

In the aforementioned equation, the term '$m$' is the first integer larger than "$a$", such as $m - 1 < \alpha < m$, and $t - \alpha$ is the interval of integration and $\Gamma(\alpha)$ is Euler's Gamma function.

**Definition 2:** The $\alpha^{th}$ order Caputo fractional derivative expression of a continuous function is formulated as follows:

$$_aD_t^\alpha \cong D^\alpha = \begin{cases} \frac{1}{\Gamma(n-\alpha)} \int_a^t \frac{f^n(\tau)}{(t-\tau)^{\alpha-n+1}} d\tau & (n-1 \leq \alpha < n) \\ \frac{d^m}{dt^n} f(t) & (\alpha = n) \end{cases} \quad (9)$$

**Definition 3:** The GL definition of order $\alpha$ is stated as follows:

507

$$^{GL}_a D_t^\alpha f(t) = \lim_{h \to 0} \frac{1}{h^\alpha} \sum_{j=0}^{[(t-\alpha)/h]} (-1)^j \binom{\alpha}{j} f(t-jh) \quad (10)$$

$$\binom{\alpha}{j} = \frac{\Gamma(\alpha+1)}{\Gamma(j+1)\Gamma(\alpha-j+1)} \quad (11)$$

In the above equation, the term '$h$' represents the time step and $\Gamma(.)$ represents the gamma function. In [15, 17], the stability of fractional order systems has been discussed in details. To approximate the fractional orders by classical integer order transfer function, an oustaloup recursive approximation algorithm is used. In [18], an oustaloup recursive approximation algorithm is discussed in detail.

### 2.4 Fractional Order Dynamics and Total Harmonic Distortion

Advantages of FO controllers have been presented in [19-21]. This section is dedicated to compare the energy decay property of the integer order with fractional order systems in the fractional range of $0<\alpha<1$.

From Fig. 4, it is evident that the release of energy in the case of fractional order systems is less and slower as compared to the integer order systems.

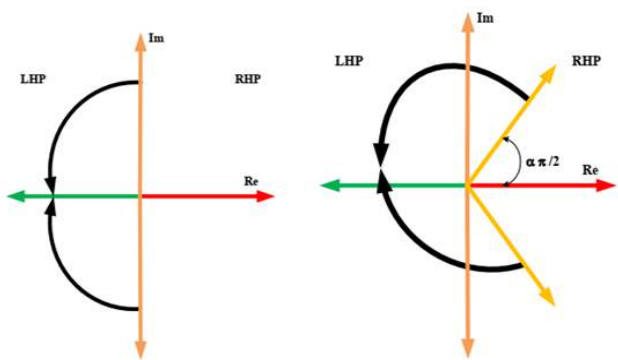

**Fig. 4:** Stability Region of:(a) Integer Order controller on the left. (b) Fractional Order on the right.

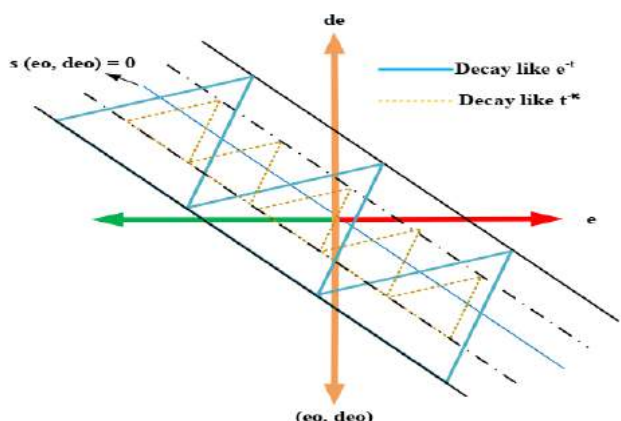

**Fig. 5:** Stability region of Integer order and Fractional order controller considering same sliding surface for both

**Remark 1:** *The slow and less intense decay property of fractional order systems can be further exploited for a low value of the total harmonic distortion (THD) of a grid-tied inverter with an L filter.*

### 2.5 State-Space Representation of PV System

The open-loop system is considered for deriving the state-space formulation of the PV system. By applying Kirchhoff's voltage and current law, the dynamics of the system are determined. The state-space equations are derived in which the $V_{pv}$ is the PVG output voltage, $I_o$ is inductor current, $V_{Dc}$ is the DC-link voltage and $I_g$ is the grid current. Via replacing state-variables $V_{pv}, I_o, V_{dc}, I_g$ by $x_1$, $x_2$, $x_3$, $x_4$ respectively we attain the following equations:

$$C_{pv}\dot{x}_1 = I_{pv} - x_2 \quad (12.a)$$

$$L_o\dot{x}_2 = x_1 - (1-\mu_1)x_3 \quad (12.b)$$

$$C_{dc}\dot{x}_3 = (1-\mu_1)x_2 - \mu_2 x_4 \quad (12.c)$$

$$L_g\dot{x}_4 = \mu_2 x_3 - V_g \quad (12.d)$$

### 3. Controller Design

As mentioned earlier in the introduction, a back-stepping controller (BSC) is chosen and derived in this section. A conventional BSC is a recursive nonlinear controller that can be used to control dynamic entities in a given model of a system. A FO controller offers more degrees of freedom as compared to the IO controller. In this regard, the slow and intense nature of a FO operator is utilized via FOBSC to reduce oscillations around dynamic entities of the system and enhance the overall power quality. In this section, a modified fractional order operator ($D^\alpha$) is introduced into conventional BSC; which is presented here as FOBSC that has an extra tuning variable ($\alpha$) to further intensify the response of derivative and integral of the system dynamics according to our desired response. As shown in Fig. 6, the controller is designed in three different loops. The first loop is designed to track the reference voltage generated by MPPT and regulate the output voltage of PVG to the point where it produces the maximum power. A FOBSC is implemented to achieve a satisfactory tracking of the reference voltage generated by MPPT. The second loop is designed to track a constant voltage reference to keep the voltage of DC-link at the desired level. A fractional-order PI (FOPI) based controller is used for this loop. The third loop is based on FOBSC. In this loop, the output of FOPI is multiplied by grid voltage to generate a sinusoidal reference in phase to the grid. FOBSC is introduced to track this reference and keep the power factor close to unity. Introducing the fractional-order parameter to back-stepping and the PI controller further reduces the THD as compared to conventional methods used till now.

### 3.1. Fractional-Order PVG Voltage Control

The objective of the first loop is to track the reference voltage ($x_1^*$) generated by MPPT and regulate the output of PVG ($x_1 = V_{pv}$) close to it. The FOBSC is implemented to track the generated reference. Initially, an error is introduced as:

$$e_1 = C_{pv}(x_1 - x_1^*) \tag{13}$$

where $x_1^*$ is the reference voltage generated by P&O based MPPT. Through the derivative of $e_1$ and using $(12.a)$, we obtain:

$$\dot{e}_1 = (I_{pv} - x_2 - C_{pv}\dot{x}_1^*) \tag{14}$$

The following Lyapunov stability candidate function is considered to prove stability:

$$V_1 = 0.5 e_1^2 \tag{15}$$

Its derivative is given by $\dot{V}_1 = e_1 \dot{e}_1$. Choosing $\dot{e}_1 = -c_1 e_1$, clearly makes the term $\dot{V}_1 = -c_1 e_1^2$ negative definite, as the square will keep the error positive while the negative sign keeps the overall result negative until the term $c_1$ is positive. Therefore, the system is globally asymptotically stable. Since at this stage, there is no control input therefore, dynamic (that exists in Eq. 14 i.e.) $x_2$ can be chosen as a virtual control input. Using $\dot{e}_1 = -c_1 e_1$ yields:

$$x_2^* = I_{pv} + c_1 e_1 - C_{pv}\dot{x}_1^* \tag{16}$$

The new virtual control variable $x_2^*$ is not a real control input, therefore, we require another error. In this error, a new operator has been introduced i.e. $D^\alpha$ which converts the equation to fractional-order. In the above-mentioned operator, $D^\alpha, \alpha$ is further tuned to improve the response of the system.

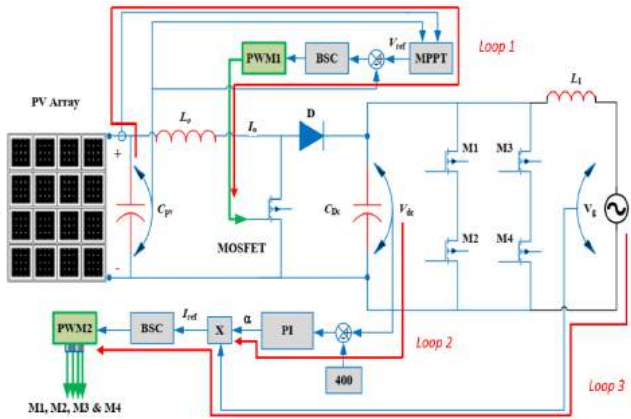

**Fig. 6:** Block diagram of the system with the considered loops

$$e_2 = L_o D^\alpha (x_2 - x_2^*) \tag{17}$$

The new $\dot{e}_1$ and $\dot{V}_1$ yields as:

$$\dot{e}_1 = -c_1 e_1 - \frac{D^{-\alpha}e_2}{L_o} \tag{18}$$

$$\dot{V}_1 = -c_1 e_1^2 - D^{-\alpha}e_1 e_2 / L_o \tag{19}$$

By time derivative of $e_2$ and using $(12.b)$ we obtain the following equation:

$$\dot{e}_2 = D^\alpha(x_1 - (1-u_1)x_3 - L_o \dot{x}_2^*) \tag{20}$$

For the stability purposes, the Lyapunov candidate function ($V_2 = 0.5 e_2^2 + V_1$) has been chosen. By time derivative of $V_2$ the equation becomes:

$$\dot{V}_2 = -c_1 e_1^2 + e_2(\frac{-D^{-\alpha}e_1}{L_o} + \dot{e}_2) \tag{21}$$

Choosing $-c_2 e_2 = \frac{-D^{-\alpha}e_1}{L_o} + \dot{e}_2$ the above equation becomes negative definite for all positive values of $c_2$.

$$\dot{V}_2 = -c_1 e_1^2 - c_2 e_2^2 \tag{22}$$

Finally, to enforce errors $e_1$ and $e_2$ to zero, we simplify the equation to find the control signal as:

$$u_1 = 1 - \frac{1}{x_3}(x_1 + c_2 D^{-\alpha}e_2 - L_o \dot{x}_2^* - D^{-2\alpha}e_1/L_o) \tag{23}$$

### 3.2. Fractional-Order PFC

The same procedure is followed in deriving FOBSC design for power factor control. Since there is only one dynamic at near grid $L_g$, therefore only one loop is required to control the grid current. The current reference chosen in this case is the multiple of a tuned constant ($\beta$) (which is the output of DC-link voltage control) and grid voltage which makes a sinusoidal signal $x_4^* = \beta * V_g$.

The current error is chosen as:

$$e_3 = L_g D^\alpha (x_4 - x_4^*) \tag{24}$$

Time-derivative of $e_3$ and putting equation $(12.d)$ in $\dot{e}_3$ we obtain:

$$\dot{e}_3 = D^\alpha(u_2 x_3 - V_g - L_g \dot{x}_4^*) \tag{25}$$

Similarly, to find stability a Lyapunov candidate function has been introduced as: $V_3 = 0.5 e_3^2$. Using the time derivative yields $\dot{V}_3 = e_3 \dot{e}_3$. Choosing $\dot{e}_3 = -c_3 e_3$ makes it negative definite.

$$\dot{V}_3 = -c_3 e_3^2 \tag{26}$$

To separate the signal of the controller the following equation has been derived:

$$u_2 = \frac{1}{x_3}(V_g + L_g \dot{x}_4^* - c_3 D^{-\alpha} e_3) \tag{27}$$

### 3.3. Fractional Order DC-link Voltage Control

One of the main challenges of PVG based inverter is the voltage across the DC-link capacitor. The power delivered by the PVG is fed to the DC-link capacitor while inverter draws the same power from it, while, at the same time a number of fluctuations are experienced by this element. At the time the PVG generate power, the capacitor voltage rises due to an increase in input power. Similarly, when inverter draws power from it, consequently voltage decreases across the DC-link capacitor. Due to significant increasing and decreasing phenomena of DC-link voltage, the output of the inverter is distorted thereby the grid power quality will be reduced. To tackle this issue, a control loop is required to keep the DC-link voltage constant. A DC-link voltage level should be greater than the grid voltage as explained in [14]. A Fractional proportional-integral (PI) based controller is applied to regulate the voltage across DC-link. The output of the PI regulator acts as a current reference ($\beta$) which is further multiplied by grid voltage to generate a sinusoidal current reference.

$$\beta = (K_p + K_i D^{-\alpha})(x_3^2 - x_3^{*2}) \quad (28)$$

Where $K_p$ is the proportional gain, $K_i$ is the integral gain and the DC-link voltage is squared to make the response of the regulator faster.

## 4. Results and Discussions

To verify the practicality of the proposed controller, a PVG having the maximum capacity of 1492W is tested in MATLAB/SIMULINK. The system specifications are given in Table 1. The IO controller is implemented according to [12]. The whole system is tested for FO controller as well as IO controller. To check the system performance three case-scenarios are studied. In the first case, the time-steps are considered from 0 to 0.5 seconds. Steady-state performance is evaluated by using ideal input parameters for PVG as $1000\ W/m^2$ Irradiance and $25\ °C$ Temperature. As it can be seen from Fig. 7a and 7b, the power output of PVG in this case is almost 100% (1492$W$) which is considered to be an ideal condition. In the second test case, the time-steps are considered to vary from 0.5 to 0.8 seconds. The system is evaluated for 80% of PVG capacity which is equal to 1202W. The irradiance and temperature inputs are $800\ W/m^2$ and 30, respectively. In the third case, the capacity of PVG is kept at 70% of the rated power which is equal to 1050.5 $W$. $700\ W/m^2$ irradiance and 35 ℃ are chosen in this case where the time-steps varies from 0.8 to 1seconds.

In Fig. 8 to 11, the study has investigated the DC side performance of the system. Figures 10 and 11 show the boost inductor current and PV output capacitor voltage. In case 1, the response of the system has been tested in an ideal weather condition. The steady state response clearly show that the FO controller has outperformed the IO controller. The current in FO controller resides around 7.3 $A$ with smaller oscillations between the highest and the lowest peak. This shows that the charging and discharging phenomena of boost inductor has been improved. This effect has clearly reduced the width of the ripple. The advantage of slow and intense nature of the FO controller has also been seen in the output voltage of PV in Fig. 11. It is evident from the PVG IV-curve that maximum output power is drawn at 203$V$ output. The voltage in Fig. 11 in FO controller is almost 203V with least oscillations between maximum charging and discharging points of PV output capacitor.

As voltage and current both have been improved, the power output in Fig. 9 is also enhanced not only in terms of oscillations but also in capacity. The power output via FO controller is above 1490.5W which is close to the ideal capacity of PVG i.e. 1492W while in IO controller the power capacity is seen to be oscillating around 1487W. Effect of FO controller is observed in P&O output reference voltage in Fig. 8. Similarly, in case 2 and 3, the reference generation in terms of FO controller compared to IO is much closer to the I-V curve. FO controller offers a reduction in peaky oscillations thereby the average power is also increased. In all the three cases, the FO controller has shown least oscillations in power while a higher average power generation capacity (considering the graph of FO controller is slightly above the IO controller graph). Additionally, the dynamic behavior of the system in case 2 and 3 is far better using FO controller as observed in Fig. 9. The lowest peak in case 2 and 3 is approximately appeared at 1158W at 0.5 sec and 975W at 0.8 sec for IO controller, where in case of FO controller it occurs at 1169W at 0.5 sec and 985 W at 0.8 sec.

**Table 1:** Specifications of the complete system

| No. | Parameters | Values |
|---|---|---|
| 1 | Module Type: Soltech 1STH-215-P | 1 |
| 2 | Parallel strings | 1 |
| 3 | Series panels | 7 |
| 4 | PV Open-circuit voltage, $V_{oc}$ | 36.4 V |
| 5 | PV Short-circuit current, $I_{sc}$ | 7.84 V |
| 6 | PV maximum power point voltage, $V_{mpp}$ | 29 V |
| 7 | PV single panel maximum output power | 215 W |
| 8 | PV total output power capability | 1492 W |
| 9 | PV output capacitor, $C_{pv}$ | 0.2 mF |
| 10 | Boost converter Inductor, $L_o$ | 100 mH |
| 11 | DC-link capacitor, $C_{dc}$ | 5 mF |
| 12 | BSC controller constant, $c_1$ | 500,000 |
| 13 | BSC controller constant, $c_2$ | 500 |
| 14 | PWM boost converter, $S_1$ | 100 KHz |
| 15 | FOBSC tuning order, $\alpha_2$ | 0.6 |
| 16 | Grid filter inductor, $L_g$ | 9 mH |
| 17 | BSC controller constant, $c_4$ | 500,000 |
| 18 | PWM inverter, $S_2$ | 10 KHz |
| 19 | FOPI tuning order, $\alpha$ | 0.95 |
| 20 | FOBSC tuning order, $\alpha_1$ | 0.875 |

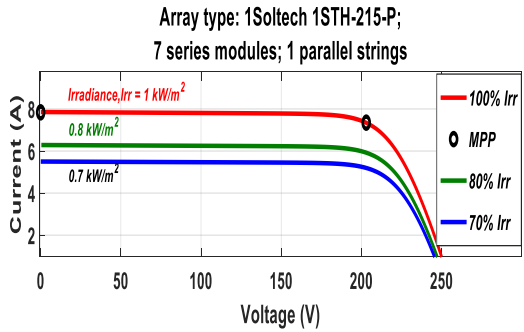

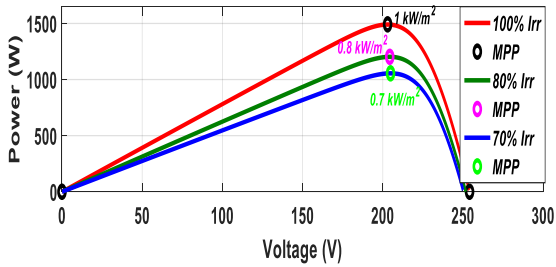

**Fig. 7a:** Effect of Irradiance on solar PV output

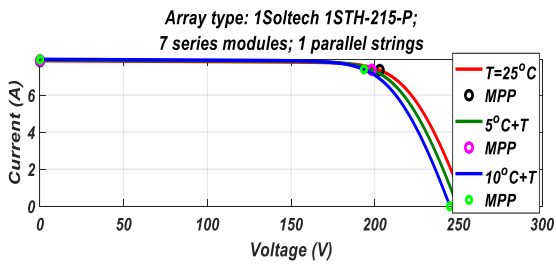

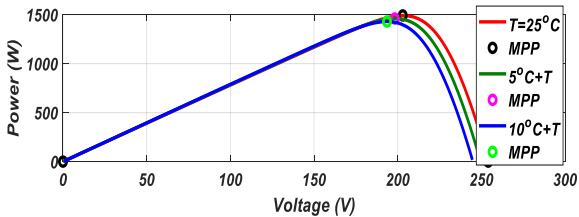

**Fig. 7b:** Effect of temperature on solar PV output

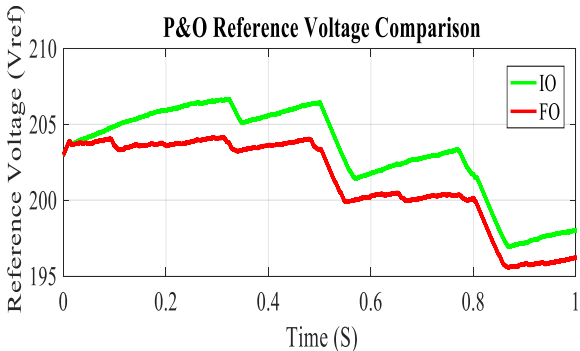

**Fig. 8:** P&O Reference Voltage – $V_{ref}$

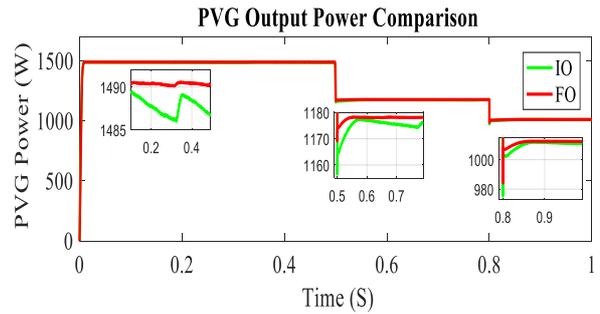

**Fig. 9:** Maximum Output Power of PVG

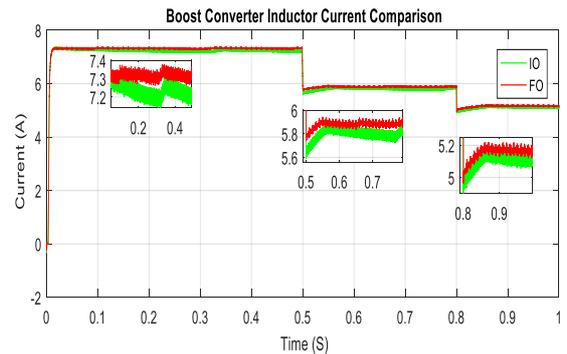

**Fig. 10:** Boost Converter Inductor Current

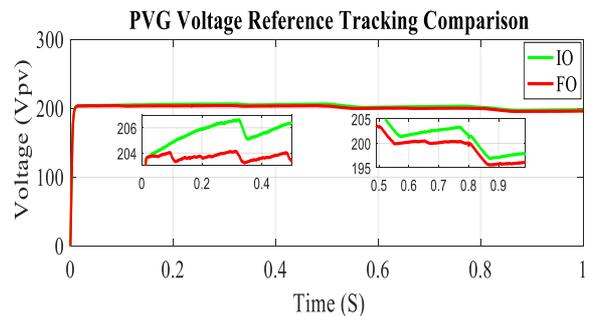

**Fig. 11:** PVG Voltage

Figures 12 to 15 show response of the AC part of the system, representing; grid current, real power, reactive power and power factor of grid respectively. Fig. 12 shows the grid current in terms of FO and IO, being compared to the reference current generation of the grid. As discussed earlier, the slow and intense nature of the FO controller offers a more fluctuation-free real power of 1405 W and reactive power of -54 VAR in case 1 of Fig. 13 and 14. The output power of the PV in terms of FO was found to be 1490.5 W. The power losses in this case for boost converter and inverter are about 5.736%. In case 2, the real and reactive power is around 1093W and -42.5VAR. Real and reactive powers in case 3 are observed to be 925W and -36VAR. In all the three cases, the real powers of the IO controller are 1399W, 1088W and 921W while reactive powers are -53VAR, -41.5VAR and -34VAR, respectively. The steady state power losses in the IO controller is approximately 5.917%. Due to less intense nature of FO controller, reduction of oscillation occurs, therefore, the power factor in Fig. 15 is also witnessed to be increased. During case 1, the power factor is nearly 0.9985 for FO controller and 0.998 for IO controller.

In case 2 and 3, the PF in IO controller is significantly lower i.e. 0.9973 while FO controller still offers a higher PF of 0.9984.

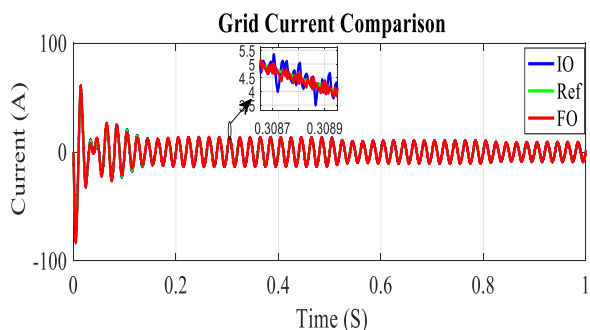

**Fig. 12:** Comparison of the grid current

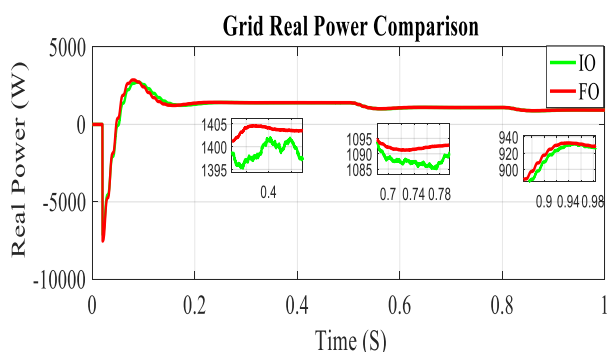

**Fig. 13:** Real Power Comparison

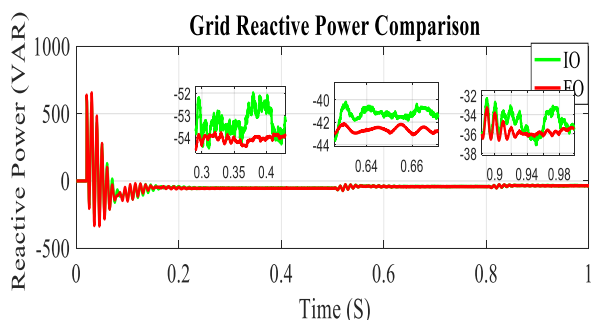

**Fig. 14:** Reactive Power Comparison

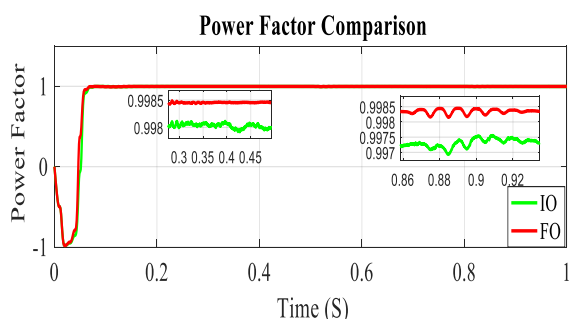

**Fig. 15:** Power Factor Comparison

The comparison of all the obtained THDs is presented in Table 2. According to the results, the FO controller has achieved considerably lower THDs in comparison to IO controller. FO controller offers THD of 4.04%, 4.19% and 4.56% for three studied cases where for the same cases IO offers 4.99%, 5.62% and 6.42%, respectively.

Table II: Specifications of the complete system

| Cases  | FO Controller | IO Controller |
|--------|---------------|---------------|
| Case 1 | 4.05%         | 5%            |
| Case 2 | 4.19%         | 5.62%         |
| Case 3 | 4.56%         | 6.42%         |

System efficiency is given in Fig. 16. As the FO controller offers smooth response in terms of oscillation thereby reducing the power losses compared to the IO controller. Therefore, the overall effect is felt on the efficiency of the overall system.

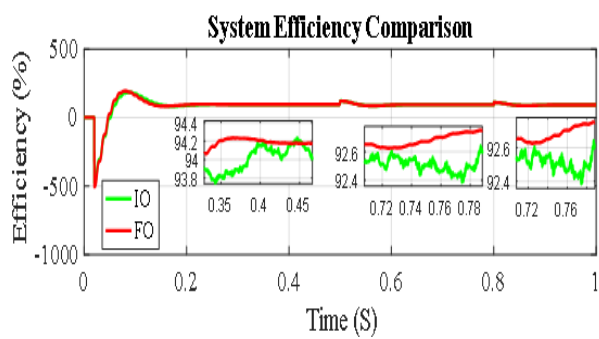

**Fig. 16:** System Efficiency

**5. Conclusion**

A non-Linear FOBSC system is developed in this study. The functionality of the proposed system is tested in MATLAB/SIMULINK. The findings of this study can be precisely presented as follows:

- The results have verified that the slow and intense nature of the FO controller can be used to enhance the power-quality of the system.

- The output power of the PVG in case of FO controller contains less oscillations than IO controller, hence the reduction of power losses is observed, which consequently improves the power generation capacity of the PVG.

- Results have shown that the FO controller develops voltages and currents with reduced oscillations in waves across the systems dynamics such as PV capacitor, boost and grid inductor.

- It is well observed in the given results that sine wave of FOBS controller is superior over the IO controller. In the conventional method, the IO controller THD is calculated as 4.99% in case 1 while in the proposed method has reduced the THD to 4.05%. According to IEEE standards, THD should be less than 5%. In two other cases, the THD of IO controller crosses the IEEE limit while the proposed controller still maintains IEEE standard.

- The proposed controller also outperforms the

conventional controller in terms of power loss. A total of 5.736% of system power losses are calculated under FO controller while in IO the system experiences about 5.917% losses.

- The power factor of the inverter is very close to unity in both controllers whereas the proposed controller offers a higher power factor compared to the IO controller. During the steady state condition, the FO controller offers the PF around 0.9985 while in IO is slightly low i.e. 0.998. During reduction in available irradiance and change in temperature, the FO offers PF of 0.9984 while the IO shows PF of about 0.9974.

**Acknowledgment**

This work has been partially supported by the Post-doctoral International Exchange Fellowship Program 2018 (Talent-Introduction Program 2018) of the P. R. China (Fund No. 207689).